\documentstyle[12pt]{article}
\newcommand{\bq}{\begin{equation}}
\newcommand{\eq}{\end{equation}}
\newcommand{\ba}{\begin{eqnarray}}
\newcommand{\ea}{\end{eqnarray}}

\newcommand{\nl }{ \nonumber  }

\newcommand{\ul}{\underline}
\newcommand{\p}{\partial}
\newcommand{\pu}{\p_\tau}
\newcommand{\pJ}{\p_J}
\newcommand{\pK}{\p_K}
\newcommand{\pj}{\p_j}
\newcommand{\pk}{\p_k}
\newcommand{\h}{\hspace{1cm}}
\newcommand{\s}{\sigma}

\newcommand{\us}{\underline\sigma}

\newcommand{\La}{\Lambda}
\newcommand{\Dj}{\bigl(\pu-\mu^{j}\pj\bigr)}
\newcommand{\Dk}{\bigl(\pu-\mu^{k}\pk\bigr)}

\begin{document}
{\bf\begin{center}
  NULL BRANES IN CURVED BACKGROUNDS
\footnote{Work supported in part by the National Science Foundation
of Bulgaria under contract $\phi-620/1996$} \vspace*{.5cm}\\ P.
Bozhilov \footnote {E-mail: bojilov@thsun1.jinr.ru; permanent
address: Dept.of Theoretical Physics,"Konstantin Preslavsky" Univ. of
Shoumen, 9700 Shoumen, Bulgaria}  \\ \it Bogoliubov Laboratory of
Theoretical Physics, \\ JINR, 141980 Dubna, Russia \vspace*{.5cm}
\end{center}}

We consider null bosonic $p$-branes in curved space-times. Some exact
solutions of the classical equations of motion and of the constraints
for the null membrane in general stationary, axially symmetrical,
four dimensional, gravity background are found.
\\

PACS number(s): 11.10.Lm, 04.25.-g, 11.27.+d


\vspace*{.5cm}

\section{\bf Introduction}
\hspace{1cm}
The null (tensionless) $p$-branes correspond to usual $p$-branes with their
tension $T_p$ taken to be zero. This relationship between null
$p$-branes and the tensionful ones may be regarded as a
generalization of the massless-massive particles correspondence.
The dynamics of the null branes in curved backgrounds is
interesting also as a generalization of the motion of null
strings in such backgrounds \cite{RZ}, \cite{Kar}, \cite{DabLar},
\cite{PorPap}, \cite{KKP}.

A model for null $p$-branes living in Friedmann-Robertson-Walker
space-time (with ${\it k}=0$) was proposed in \cite{RZh}. The
motion equations were solved and it was shown there that an ideal
fluid of null $p$-branes may be considered as a source of gravity.

 Here we investigate the classical evolution of the null branes in
a curved background. In Sec. 2 we give the corresponding
Lagrangian formulation.  In Sec. 3 the BRST-BFV approach in its
Hamiltonian version is applied to the considered dynamical system.
In Sec. 4 some exact solutions of the equations of motion and of
the constraints for the null membrane in general stationary,
axially symmetrical, four dimensional, gravity background are
found. The examples of Minkowski, de Sitter, Schwarzschild,
Taub-NUT and Kerr space-times are considered in Sec. 5. Sec. 6 is
devoted to comments and conclusions.

\section{\bf Lagrangian formulation}
\hspace{1cm}
The action for the bosonic null $p$-brane in a D-dimensional curved
space-time with metric tensor $g_{\mu\nu}(x)$ can be written in the
form:
\ba\label{a}
S=\int d^{p+1}\xi L \h,\h L=V^JV^K\pJ x^\mu\pK
x^\nu g_{\mu\nu}(x),
\\ \nl
\pJ=\p/\p\xi^J, \h \xi^J=(\xi^0,\xi^j)=(\tau,\s^j),
\\ \nl
J,K=0,1,...,p \h,\h j,k=1,...,p \h,\h \mu,\nu=0,1,...,D-1.
\ea
It is an obvious generalization of the flat space-time action
given in \cite{CQG}.

To prove the invariance of the action under infinitesimal
diffeomorphisms on the world volume (reparametrizations),
we first write down the corresponding
transformation law for the (r,s)-type tensor density of weight $a$
\ba\nl
\delta_{\varepsilon}T^{J_1...J_r}_{K_1...K_s}[a]&=&
L_{\varepsilon}T^{J_1...J_r}_{K_1...K_s}[a]=
\varepsilon^L\p_L T^{J_1...J_r}_{K_1...K_s}[a]\\
\label{diff}
&+&
T^{J_1...J_r}_{KK_2...K_s}[a]\p_{K_1}\varepsilon^K+...+
T^{J_1...J_r}_{K_1...K_{s-1}K}[a]\p_{K_s}\varepsilon^K \\ \nl
&-&
T^{JJ_2...J_r}_{K_1...K_s}[a]\p_J\varepsilon^{J_1}-...-
T^{J_1...J_{r-1}J}_{K_1...K_s}[a]\p_J\varepsilon^{J_r} \\ \nl
&+&
aT^{J_1...J_r}_{K_1...K_s}[a]\p_L\varepsilon^L ,
\ea
where $L_\varepsilon$ is the Lie derivative along the vector field
$\varepsilon$. Using (\ref{diff}), one verifies that if
$x^\mu(\xi)$, $g_{\mu\nu}(\xi)$ are world-volume scalars
($a=0$) and $V^J(\xi)$ is a world-volume (1,0)-type tensor density
of weight $a=1/2$, then $\pJ x^\nu$ is a (0,1)-type tensor,
$\pJ x^\mu \pK x^\nu g_{\mu \nu}$ is a (0,2)-type tensor and $L$ is
a scalar density of weight $a=1$. Therefore,
\ba\nl
\delta_{\varepsilon}S=\int d^{p+1}\xi\p_J\bigl ( \varepsilon^J L\bigr )
\ea
and this variation vanishes under suitable boundary conditions.

The equations of motion following from (\ref{a}) are:
\ba\nl
\pJ\Bigl (V^J V^K \pK x^{\lambda}\Bigr ) +
\Gamma^{\lambda}_{\mu\nu} V^J V^K \pJ x^{\mu}\pK x^{\nu} = 0 ,\\
\nl
V^J \pJ x^{\mu}\pK x^{\nu} g_{\mu\nu}(x) = 0 ,
\ea
where $\Gamma^{\lambda}_{\mu\nu}$ is the connection
compatible with the metric $g_{\mu\nu}(x)$:
\ba\nl
\Gamma^{\lambda}_{\mu\nu}=\frac{1}{2}g^{\lambda\rho}\bigl(
\p_{\mu}g_{\nu\rho}+\p_{\nu}g_{\mu\rho}-\p_{\rho}g_{\mu\nu}\bigr) .
\ea

For the transition to Hamiltonian picture it is convenient to
rewrite the Lagrangian density (\ref{a}) in the form
($\pu=\p/\p\tau, \pj=\p/\p\s^j$):
\ba\label{L}
L=\frac{1}{4\mu^0} g_{\mu\nu}(x)\bigl (\pu-\mu^j\pj\bigr )x^\mu
\bigl (\pu-\mu^k\pk\bigr )x^\nu ,
\ea
where
\ba\nl
V^J=\bigl(V^0,V^j\bigr)=\Biggl(-\frac{1}{2\sqrt{\mu^0}},
\frac{\mu^j}{2\sqrt{\mu^0}}\Biggr) .
\ea
Now the equation of motion for $x^\nu$ takes the form:
\ba\label{eqx}
\pu\Bigl [\frac{1}{2\mu^0}\bigl (\pu-\mu^k\pk\bigr )x^{\lambda}\Bigr ]
-\pj\Bigl [\frac{\mu^{j}}{2\mu^{0}}\bigl (\pu-\mu^k\pk\bigr )x^{\lambda}
\Bigr ] \\
\nl
+ \frac{1}{2\mu^0}\Gamma^{\lambda}_{\mu\nu}
\bigl (\pu-\mu^j\pj\bigr )x^\mu \bigl (\pu-\mu^k\pk\bigr )x^\nu = 0 .
\ea
The equations of motion for the Lagrange multipliers $\mu^{0}$ and
$\mu^{j}$ which follow from (\ref{L}) give the constraints :
\ba\label{Tx1}
g_{\mu\nu}(x)\bigl (\pu-\mu^j\pj\bigr )x^\mu
\bigl (\pu-\mu^k\pk\bigr )x^\nu = 0 , \\
\nl
g_{\mu\nu}(x)\bigl (\pu-\mu^k\pk\bigr )x^\mu \pj x^\nu = 0 .
\ea
In terms of $x^\nu$ and the conjugated momentum $p_\nu$ they read:
\ba\label{Tpx}
T_0=g^{\mu\nu}(x)p_\mu p_\nu = 0 \h,\h T_j=p_\nu\pj x^\nu = 0 .
\ea

\section{\bf Hamiltonian formulation}
\hspace{1cm}
The Hamiltonian which corresponds to the Lagrangian density
(\ref{L}) is a linear combination of the constraints (\ref{Tpx}) :
\ba\nl
H_0=\int d^p\sigma\bigl (\mu^0 T_0+\mu^j T_j \bigr ) .
\ea
They satisfy the following (equal $\tau$) Poisson bracket algebra
\ba \nl
\{T_0(\ul \sigma_1),T_0(\ul \sigma_2)\}&=&0,
\\ \label {CA}
\{T_0(\ul \sigma_1),T_{j}(\ul \sigma_2)\}&=& [T_0(\ul \sigma_1) + T_0(\ul
\sigma_2)] \p_j \delta^p (\ul \sigma_1 - \ul \sigma_2) ,
\\ \nl
\{T_{j}(\ul \sigma_1),T_{k}(\ul \sigma_2)\}&=&
[\delta_{j}^{l}T_{k}(\ul \sigma_1) + \delta_{k}^{l}T_{j}(\ul
\sigma_2)]\p_l\delta^p(\ul \sigma_1-\ul \sigma_2) ,\\
\nl
\us=(\s^1,...,\s^p) .
\ea
The equalities (\ref{CA}) show that the constraint algebra is the
same for flat and for curved backgrounds. Having in mind the above
algebra, one can construct the corresponding BRST charge $\Omega$
\cite{FF} (*=complex conjugation)
\ba \label{O}
\Omega = \Omega^{min}+\pi_J \bar {\cal P}^J , \h
\{\Omega,\Omega\} = 0 , \h \Omega^* = \Omega .
\ea
$\Omega^{min}$ in (\ref{O}) can be written as \cite{MPL}
\ba\nl
\Omega^{min}&=&\int d^p\s\{T_0\eta^0+T_j\eta^j+
{\cal P}_0 [(\p_j\eta^j)\eta^0 + (\p_j\eta^0)\eta^j ] +
{\cal P}_k(\p_j\eta^k)\eta^j \} ,
\ea
and can be represented also in the form
\ba\nl
\Omega^{min}=\int d^p\s\bigl [\bigl (T_0+
\frac{1}{2}T_0^{gh}\bigr )\eta^0
+\bigl (T_j+\frac{1}{2}T_j^{gh}\bigr )\eta^j \bigr ]
+\int d^p\s\p_j\Bigl (\frac{1}{2}{\cal P}_k\eta^k\eta^j \Bigr ) .
\ea
Here a superscript $gh$ is used for the ghost part of the total gauge
generators
\ba\nl
\nl
T_J^{tot}=\{\Omega,{\cal P}_J\}=\{\Omega^{min},{\cal P}_J\}=
T_J+T_J^{gh} .
\ea
We recall that the Poisson bracket algebras of $T_J^{tot}$ and
$T_J$ coincide for first rank systems which is the case under
consideration.  The manifest expressions for $T_J^{gh}$ are:
\ba\nl
T_0^{gh}&=&2{\cal P}_0\p_j\eta^j+\bigl (\p_j{\cal P}_0\bigr )\eta^j ,
\\ \nl
T_j^{gh}&=&2{\cal P}_0\p_j\eta^0+\bigl (\p_j{\cal P}_0\bigr )\eta^0+
{\cal P}_j\p_k\eta^k+{\cal P}_k\p_j\eta^k+
\bigl (\p_k{\cal P}_j\bigr )\eta^k .
\ea
Up to now, we introduced canonically conjugated ghosts
$\bigl (\eta^J,{\cal P}_J\bigr )$,
$\bigl (\bar \eta_J,\bar {\cal P}^J\bigr)$ and momenta $\pi_J$ for
the Lagrange multipliers $\mu^J$ in the Hamiltonian. They have
Poisson brackets and Grassmann parity as follows ($\epsilon_J$ is
the Grassmann parity of the corresponding constraint):
\ba\nl
\bigl \{\eta^J,{\cal P}_K\bigr \}&=&\delta^J_K ,
\h
\epsilon (\eta^J)=\epsilon ({\cal P}_J)=\epsilon_J + 1 , \\
\nl
\bigl \{\bar \eta_J,\bar {\cal P}^K \bigr \}&=&-(-1)^{\epsilon_J\epsilon_K}
\delta^K_J ,
\h
\epsilon (\bar\eta_J)=\epsilon (\bar {\cal P}^J)=\epsilon_J + 1 ,
\\ \nl
\bigl \{\mu^J,\pi_K\bigr \}&=&\delta^J_K ,
\h
\epsilon (\mu^J)=\epsilon (\pi_J)=\epsilon_J .
\ea

The BRST-invariant Hamiltonian is
\ba\label{H}
H_{\tilde \chi}=H^{min}+\bigl \{\tilde \chi,\Omega\bigr \}=
\bigl \{\tilde \chi,\Omega\bigr \} ,
\ea
because from $H_{canonical}=0$ it follows $H^{min}=0$. In this
formula $\tilde \chi$ stands for the gauge fixing fermion
$(\tilde \chi^* = -\tilde \chi)$. We use the following representation
for the latter
\ba\nl
\tilde \chi=\chi^{min}+\bar\eta_J(\chi^J+\frac{1}{2}\rho_{(J)}\pi^J) ,
\h
\chi^{min}=\mu^J{\cal P}_J ,
\ea
where $\rho_{(J)}$ are scalar parameters and we have separated the
$\pi^J$-dependence from $\chi^J$. If we adopt that $\chi^J$ does
not depend on the ghosts $(\eta^J,{\cal P}_J)$ and
$(\bar\eta_J,\bar {\cal P}^J)$, the Hamiltonian $H_{\tilde\chi}$
from (\ref{H}) takes the form
\ba
\label{r}
H_{\tilde\chi}&=&H_{\chi}^{min}+{\cal P}_J \bar {\cal P}^J -
\pi_J(\chi^J+\frac{1}{2}\rho_{(J)}\pi^J)+ \\
\nl
&+&\bar\eta_J \bigl \{\chi^J,T_K\bigr \}\eta^K ,
\ea
where
\ba\nl
H_{\chi}^{min}=\bigl \{\chi^{min},\Omega^{min}\bigr \} .
\ea

One can use the representation (\ref{r}) for $H_{\tilde\chi}$ to obtain the
corresponding BRST invariant Lagrangian density
\ba\nl
L_{\tilde\chi}=L+L_{GH}+L_{GF} .
\ea
Here $L$ is given in (\ref{L}), $L_{GH}$ stands for the ghost part
and $L_{GF}$ - for the gauge fixing part of the Lagrangian density.
The manifest expressions for $L_{GH}$ and $L_{GF}$ are \cite{MPL}:
\ba
\nl
L_{GH}=-\p_\tau\bar{\eta_0}\p_\tau\eta^0-\p_\tau\bar{\eta_j}
\p_\tau\eta^j+\mu^0[2\pu\bar{\eta_0}\pj\eta^j+
(\pj\pu\bar{\eta_0})\eta^j]
\\
\nl
+\mu^j[2\pu\bar{\eta_0}\pj\eta^0+
(\pj\pu\bar{\eta_0})\eta^0+\pu\bar{\eta_k}\pj\eta^k+
\pu\bar{\eta_j}\p_k\eta^k+
(\p_k\pu\bar{\eta_j})\eta^k]
\\
\nl
+\int d^p\sigma'\{\bar{\eta_0}(\underline{\sigma'})
[\{T_0,\chi^0(\underline{\sigma'})\}\eta^0
+\{T_j,\chi^0(\underline{\sigma'})\}\eta^j]
\\
\nl
+\bar{\eta_j}(\underline{\sigma'})[\{T_0,\chi^j(\underline{\sigma'})\}
\eta^0+\{T_k,\chi^j(\underline{\sigma'})\}\eta^k]\} , \\
\nl
L_{GF}=\frac{1}{2\rho_{(0)}}(\p_\tau \mu^0-\chi^0)(\p_\tau
\mu_0-\chi_0)+ \frac{1}{2\rho_{(j)}}(\p_\tau \mu^j-\chi^j) (\p_\tau
\mu_j-\chi_j) .
\ea

If one does not intend to pass to the Lagrangian
formalism, one may restrict oneself to
the minimal sector $\bigl (\Omega^{min},\chi^{min},H_\chi^{min}\bigr )$.
In particular, this means that Lagrange multipliers are not considered as
dynamical variables anymore.
With this particular gauge choice, $H_\chi^{min}$
is a linear combination of the total constraints
\ba\nl
H_\chi^{min}=\int d^p\s\Bigl [\La^0 T_0^{tot}(\us)+\La^j
T_j^{tot}(\us)\Bigr ] ,
\ea
and we can treat here the Lagrange multipliers $\La^0,\La^{j}$
as constants. Of course, this does not fix the gauge completely.

\section{\bf Null membranes in D=4}
\hspace{1cm}
Here we confine ourselves to the case of membranes moving in a
four dimensional, stationary, axially symmetrical, gravity
background of the type
\ba\label{gm}
ds^2&=&g_{00}(dx^0)^2+g_{11}(dx^1)^2+g_{22}(dx^2)^2+
g_{33}(dx^3)^2+2g_{03}dx^0 dx^3 , \\
\nl
g_{\mu\nu}&=&g_{\mu\nu}(x^1,x^2) .
\ea
We will work in the gauge $\mu^0, \mu^j = constants $, in which the
equations of motion (\ref{eqx}) and constraints (\ref{Tx1})
for the membrane $(j,k=1,2)$ have the form:
\ba\label{eqxf}
\bigl(\pu-\mu^j\pj\bigr)\bigl (\pu-\mu^k\pk\bigr )x^{\lambda}
+ \Gamma^{\lambda}_{\mu\nu}
\bigl (\pu-\mu^j\pj\bigr )x^\mu \bigl (\pu-\mu^k\pk\bigr )x^\nu = 0 .
\\ \label{cf1}
g_{\mu\nu}(x)\bigl (\pu-\mu^j\pj\bigr )x^\mu
\bigl (\pu-\mu^k\pk\bigr )x^\nu = 0 , \\
\nl
g_{\mu\nu}(x)\bigl (\pu-\mu^k\pk\bigr )x^\mu \pj x^\nu = 0 .
\ea
To establish the correspondence with the null geodesics we note
that if we introduce the quantities
\ba\label{unu}
u^{\nu}(x)=\bigl(\pu-\mu^{j}\pj\bigr)x^{\nu} ,
\ea
the equations of motion (\ref{eqxf}) can be rewritten as
\ba\nl
u^{\nu}\bigl(\p_{\nu}u^{\lambda}+\Gamma^{\lambda}_{\mu\nu}u^{\mu}\bigr)=0 .
\ea
Then it follows from here that $u^2$ do not depend on $x^{\nu}$.
In this notations, the constraints are:
\ba\nl
g_{\mu\nu}u^{\mu}u^{\nu} = 0 \h,\h
g_{\mu\nu}u^{\mu}\pj x^{\nu} = 0 .
\ea

Taking into account the metric (\ref{gm}), one can write the
equations of motion (\ref{eqxf}) and the constraints (\ref{cf1})
in the form:
\ba\nl
\Dj u^0+2\Bigl(\Gamma^0_{01}u^0+\Gamma^0_{13}u^3\Bigr)\Dj x^1 +
\\ \nl
+2\Bigl(\Gamma^0_{02}u^0+\Gamma^0_{23}u^3\Bigr)\Dj x^2 = 0 ,\\
\nl
\Dj\Dk x^1+\Gamma^1_{11}\Dj x^1\Dk x^1+
\\ \nl
+2\Gamma^1_{12}\Dj x^1\Dk x^2 +\Gamma^1_{22}\Dj x^2\Dk x^2+
\\ \nl
+\Gamma^1_{00}(u^0)^2+ 2\Gamma^1_{03}u^0 u^3+\Gamma^1_{33}(u^3)^2 = 0 ,
\\ \label{l2}
\Dj\Dk x^2+\Gamma^2_{11}\Dj x^1\Dk x^1+
\\ \nl
+2\Gamma^2_{12}\Dj x^1\Dk x^2+\Gamma^2_{22}\Dj x^2\Dk x^2+
\\ \nl
+\Gamma^2_{00}(u^0)^2+2\Gamma^2_{03}u^0 u^3+\Gamma^2_{33}(u^3)^2 = 0 ,
\\ \nl
\Dj u^3+2\Bigl(\Gamma^3_{01}u^0+\Gamma^3_{13}u^3\Bigr)\Dj x^1 +
\\ \nl
+2\Bigl(\Gamma^3_{02}u^0+\Gamma^3_{23}u^3\Bigr)\Dj x^2 = 0 ,\\
\nl
g_{11}\Dj x^1\Dk x^1+g_{22}\Dj x^2\Dk x^2+
\\ \nl
+g_{00}(u^0)^2+2g_{03}u^0 u^3+g_{33}(u^3)^2 = 0 ,\\
\nl
g_{11}\Dk x^1\pj x^1+g_{22}\Dk x^2\pj x^2 +
\\ \nl
+\bigl(g_{00}\pj x^0+g_{03}\pj x^3\bigr)u^0 +
\bigl(g_{03}\pj x^0+g_{33}\pj x^3\bigr)u^3 = 0 ,
\ea
where the notation introduced in (\ref{unu}) is used. To simplify
these equations, we make the ansatz
\ba\nl
x^0(\tau,\us)&=&f^0(z^1,z^2)+t(\tau) ,
\\ \label{az}
x^1(\tau,\us)&=&r(\tau) \h,\h
x^2(\tau,\us)=\theta(\tau) ,
\\ \nl
x^3(\tau,\us)&=&f^3(z^1,z^2)+\varphi(\tau) ,
\\ \nl
z^j&=&\mu^j \tau + \sigma^j ,
\ea
where $f^0, f^3$ are arbitrary functions of their arguments.

After substituting (\ref{az}) in (\ref{l2}), we receive
(the dot is used for differentiation with respect to the affine
parameter $\tau$):
\ba\label{l01}
\dot u^0+\Bigl[\left(g^{00}\p_1 g_{00}+g^{03}\p_1 g_{03}\right)u^0
+\left(g^{00}\p_1 g_{03}+g^{03}\p_1 g_{33}\right)u^3\Bigr]\dot r
\\ \nl
+\Bigl[\left(g^{00}\p_2 g_{00}+g^{03}\p_2 g_{03}\right)u^0
+\left(g^{00}\p_2 g_{03}+g^{03}\p_2
g_{33}\right)u^3\Bigr]\dot\theta = 0 ,
\\ \label{l11}
2g_{11}\ddot r+\p_1 g_{11}\dot r^2+2\p_2 g_{11}\dot r\dot\theta
-\p_1 g_{22}\dot\theta^2
\\ \nl
-\bigl[\p_1 g_{00}(u^0)^2+2\p_1 g_{03}u^0 u^3
+\p_1 g_{33}(u^3)^2\bigr]=0 ,
\\ \label{l21}
2g_{22}\ddot\theta+\p_2 g_{22}\dot\theta^2+2\p_1 g_{22}\dot
r\dot\theta-\p_2 g_{11}\dot r^2
\\ \nl
-\bigl[\p_2 g_{00}(u^0)^2+2\p_2 g_{03}u^0 u^3
+\p_2 g_{33}(u^3)^2\bigr]=0 ,
\\ \label{l31}
\dot u^3+\Bigl[\left(g^{33}\p_1 g_{03}+g^{03}\p_1 g_{00}\right)u^0
+\left(g^{33}\p_1 g_{33}+g^{03}\p_1 g_{03}\right)u^3\Bigr]\dot r
\\ \nl
+\Bigl[\left(g^{33}\p_2 g_{03}+g^{03}\p_2 g_{00}\right)u^0
+\left(g^{33}\p_2 g_{33}+g^{03}\p_2
g_{03}\right)u^3\Bigr]\dot\theta = 0 ,
\\ \label{c11}
g_{11}\dot r^2+g_{22}\dot\theta^2+g_{00}(u^0)^2
+2g_{03}u^0 u^3+g_{33}(u^3)^2 = 0 ,
\\ \label{c21}
\bigl(g_{00}\pj f^0+g_{03}\pj f^3\bigr)u^0
+\bigl(g_{03}\pj f^0+g_{33}\pj f^3\bigr)u^3 = 0 .
\ea
If we choose
\ba\nl
f^0(z^1,z^2)=f^0(w) \h,\h f^3(z^1,z^2)=f^3(w) ,
\ea
where $w=w(z^1,z^2)$ is an arbitrary function of $z^1$ and $z^2$,
then the system of equations (\ref{c21}) reduces to the single
equation
\ba\label{c22}
\Bigl(g_{00}\frac{df^0}{dw}+g_{03}\frac{df^3}{dw}\Bigr)u^0
+\Bigl(g_{03}\frac{df^0}{dw}+g_{33}\frac{df^3}{dw}\Bigr)u^3=0 .
\ea

To be able to separate the variables $u^0, u^3$ in the system of
differential equations (\ref{l01}), (\ref{l31}) with the help of
(\ref{c22}), we impose the following condition on $f^0(w)$ and
$f^3(w)$ \ba\nl f^0(w) = C^0 f[w(z^1,z^2)] \h,\h f^3(w) = C^3
f[w(z^1,z^2)] , \ea where $C^0, C^3$ are constants, and $f(w)$ is an
arbitrary function of $w$. Then the solution of (\ref{l01}),
(\ref{l31}) and (\ref{c22}) is \cite{PorPap} ($C_1 = const $): \ba\nl
u^0(\tau)=-C_1\left(C^0 g_{03}+C^3 g_{33}\right)\exp(-H) ,
\\ \label{s032}
u^3(\tau)=+C_1\left(C^0 g_{00}+C^3 g_{03}\right)\exp(-H) ,
\\ \nl
H = \int\Bigl(g^{00}dg_{00}+2g^{03}dg_{03}+g^{33}dg_{33}\Bigr) .
\ea
The condition for the compatibility of (\ref{s032}) with
(\ref{l11}), (\ref{l21}) and (\ref{c11}) is:
\ba\nl
u^0(\tau)&=&-C_1\left(C^0 g_{03}+C^3 g_{33}\right)h^{-1}
\\ \nl
&=&-C_1\left(C^3 g^{00}-C^0 g^{03}\right)=\dot t(\tau) ,
\\ \label{su3}
u^3(\tau)&=&+C_1\left(C^0 g_{00}+C^3 g_{03}\right)h^{-1}
\\ \nl
&=&-C_1\left(C^3 g^{03}-C^0 g^{33}\right)=\dot\varphi (\tau) ,
\\ \nl
h&=&g_{00}g_{33}-g_{03}^2 .
\ea

On the other hand, from (\ref{l21}) and (\ref{c11}) one has:
\ba\label{r.2}
\dot r^2&=&-g^{11}\Bigl[C_1^2\frac{G}{h}+g^{22}\left(g_{22}^2
\dot\theta^2\right)\Bigr]
\\ \nl
&=&g^{11}\Bigl\{C_1^2\bigl[2C^0 C^3 g^{03}-(C^3)^2 g^{00}-(C^0)^2
g^{33}\bigr]-g^{22}\bigl(g_{22}^2\dot\theta^2\bigr)\Bigr\} ,
\\ \label{t.2}
g_{22}^2\dot\theta^2&=&C_2+C_1^2\int\limits^{\theta}d\theta h^{-2}
\Bigl[g_{22}G\frac{\p h}{\p\theta}-h\frac{\p g_{22}G}{\p\theta}\Bigr],
\\ \nl
G&=&(C^0)^2 g_{00}+2C^0 C^3 g_{03}+(C^3)^2 g_{33}.
\ea
In obtaining (\ref{t.2}), we have used
the gauge freedom in the metric (\ref{gm}),
to impose the condition \cite{Chandra}:
\ba\nl
\p_2\Biggl(\frac{g_{22}}{g_{11}}\Biggr) = 0 .
\ea

As a final result we have
\ba\nl
x^0 &=& C^0 f[w(z^1,z^2)] + t(\tau), \\ \nl
x^1 &=& r(\tau), \\ \nl
x^2 &=& \theta (\tau), \\ \nl
x^3 &=& C^3 f[w(z^1,z^2)] + \varphi (\tau) ,
\ea
where $\dot t(\tau), \dot r(\tau), \dot\theta (\tau),
\dot\varphi (\tau)$ are given by (\ref{su3}), (\ref{r.2}),
and (\ref{t.2}).

In the particular case when $x^2=\theta=\theta_0=const$, one can
integrate to obtain the following exact solution of the equations of
motion and constraints for the null membrane in the gravity
background (\ref{gm}): \ba\nl x^0 (\tau,\s^1,\s^2)&=&C^0
f[w(z^1,z^2)] + t_0 \\ \nl &\pm&\int\limits_{r_0}^{r}dr\left(C^3
g^{00}-C^0 g^{03}\right)W^{-1/2} ,
\\ \label{GSC}
x^3 (\tau,\s^1,\s^2)&=&C^3 f[w(z^1,z^2)] + \varphi_{0}
\\ \nl
&\pm&\int\limits_{r_0}^{r}dr\left(C^3 g^{03}-C^0
g^{33}\right)W^{-1/2} ,
\\ \nl
C_1 (\tau -\tau_{0})&=& \pm\int\limits_{r_0}^{r}drW^{-1/2} ,\\ \nl
W&=&g^{11}\left[2C^0 C^3 g^{03}-\left(C^3\right)^2
g^{00}-\left(C^0\right)^2 g^{33}\right] ,
\\ \nl
t_0, r_0, \varphi_{0}, \tau_{0} &-& constants.
\ea

\section{\bf Examples}
\hspace{1cm} Here we give some examples of solutions of the type
received in the previous section. To begin with, let us start with
the simplest case of {\it Minkowski space-time}. The metric is \ba\nl
g_{00}=-1,\h g_{11}=1,\h g_{22}=r^2,\h g_{33}=r^2 \sin^2\theta , \ea
and equalities (\ref{su3}), (\ref{r.2}), (\ref{t.2}) take the form:
\ba\nl \dot t&=&C_1 C^3 ,\\ \nl \dot r^2&=&(C_1 C^3)^2 -
\frac{C_2}{r^2} ,\\ \nl r^4 \dot\theta^2&=&C_2-\frac{(C_1
C^0)^2}{\sin^2\theta} ,
\\ \nl
\dot\varphi&=&\frac{C_1 C^0}{r^2 \sin^2\theta} . \ea When
$\theta=\theta_0=const$, the solution (\ref{GSC}) is: \ba\nl x^0
(\tau,\s^1,\s^2)&=&C^0 f[w(z^1,z^2)] + t_0 \mp C^3
\int\limits_{r_0}^{r}\frac{dr} {\bigl [(C^3)^2-(C^0)^2
r^{-2}\sin^{-2}\theta_0\bigr ]^{1/2}} ,
\\ \nl
x^3 (\tau,\s^1,\s^2)&=&C^3 f[w(z^1,z^2)] + \varphi_{0}
\mp\frac{C^0}{\sin^2\theta_0}\int\limits_{r_0}^{r}\frac{dr} {r^2\bigl
[(C^3)^2-(C^0)^2 r^{-2}\sin^{-2}\theta_0\bigr ]^{1/2}} ,
\\ \nl
C_1(\tau - \tau_{0}) &=& \pm\int\limits_{r_0}^{r}\frac{dr}{\bigl
[(C^3)^2 -(C^0)^2 r^{-2}\sin^{-2}\theta_{0}\bigr ]^{1/2}} . \ea

Our next example is the {\it de Sitter space-time}. We take the
metric in the form \ba\nl g_{00}=-\left(1-kr^2\right),
g_{11}=\left(1-kr^2\right)^{-1}, g_{22}=r^2, g_{33}=r^2 \sin^2\theta
, \ea where $k$ is the constant curvature. Now we have \ba\nl \dot
t&=&\frac{C_1 C^3}{1-kr^2} ,\\ \nl \dot r^2&=&(C_1 C^3)^2 + C_2
(k-r^{-2}) ,
\\ \nl
r^4 \dot\theta^2&=&C_2-\frac{(C_1 C^0)^2}{\sin^2\theta} ,
\\ \nl
\dot\varphi&=&\frac{C_1 C^0}{r^2 \sin^2\theta} , \ea and the
corresponding solution (\ref{GSC}) is: \ba\nl x^0
(\tau,\s^1,\s^2)&=&C^0 f[w(z^1,z^2)] + t_0
\\ \nl
&\mp& C^3 \int\limits_{r_0}^{r}\frac{dr}{(1-kr^2) \bigl
[(C^3)^2+(C^0)^2 (k-r^{-2})\sin^{-2}\theta_0\bigr ]^{1/2}} ,
\\ \nl
x^3 (\tau,\s^1,\s^2)&=&C^3 f[w(z^1,z^2)] + \varphi_{0}
\\ \nl
&\mp&\frac{C^0}{\sin^2\theta_0}\int\limits_{r_0}^{r}\frac{dr}
{r^2\bigl[(C^3)^2+(C^0)^2 (k-r^{-2})\sin^{-2}\theta_0\bigr]^{1/2}} ,
\\ \nl
C_1(\tau - \tau_{0}) &=& \pm\int\limits_{r_0}^{r}\frac{dr}{\bigl
[(C^3)^2 +(C^0)^2 (k-r^{-2})\sin^{-2}\theta_0\bigr ]^{1/2}} . \ea

Now let us turn to the case of {\it Schwarzschild space-time}.
The corresponding metric may be written as
\ba\nl
g_{00}&=&-(1-2Mr^{-1}) \h,\h g_{11}=(1-2Mr^{-1})^{-1} ,
\\ \nl
g_{22}&=&r^2 \h,\h g_{33} = r^2 \sin^2\theta , \ea where $M$ is the
Schwarzschild mass. The equalities (\ref{su3}), (\ref{r.2}) and
(\ref{t.2}) read \ba\nl \dot t&=&\frac{C_1 C^3}{1-2Mr^{-1}} ,\\ \nl
\dot r^2&=&(C_1 C^3)^2 -\frac{C_2}{r^2}(1-2Mr^{-1}) ,
\\ \label{S}
r^4 \dot\theta^2&=&C_2-\frac{(C_1 C^0)^2}{\sin^2\theta} ,
\\ \nl
\dot\varphi&=&\frac{C_1 C^0}{r^2 \sin^2\theta} . \ea When
$\theta=\theta_0=const$, one obtains from (\ref{GSC}) \ba\nl x^0
(\tau,\s^1,\s^2)&=&C^0 f[w(z^1,z^2)] + t_0
\\ \nl
&\mp& C^3\int\limits_{r_0}^{r}\frac{dr}{(1-2Mr^{-1}) \bigl [(C^3)^2
-(C^0)^2 r^{-2}(1-2Mr^{-1})\sin^{-2}\theta_0]^{1/2}} ,
\\ \nl
x^3 (\tau,\s^1,\s^2)&=&C^3 f[w(z^1,z^2)] + \varphi_{0}
\\ \nl
&\mp&\frac{C^0}{\sin^2\theta_0}\int\limits_{r_0}^{r}\frac{dr}
{r^2\bigl[(C^3)^2 -(C^0)^2
r^{-2}(1-2Mr^{-1})\sin^{-2}\theta_0\bigr]^{1/2}} ,
\\ \nl
C_1(\tau - \tau_{0}) &=& \pm\int\limits_{r_0}^{r}\frac{dr}{\bigl
[(C^3)^2-(C^0)^2 r^{-2} (1-2Mr^{-1})\sin^{-2}\theta_0\bigr ]^{1/2}} .
\ea

For the {\it Taub-NUT space-time} we take the metric as \ba\nl
g_{00}=-\frac{\delta}{R^2}\h,\h g_{11}=\frac{R^2}{\delta}\h,\h
g_{22}=R^2 ,\\ \nl
g_{33}=R^2\sin^2\theta-4l^2\frac{\delta\cos^2\theta}{R^2}\h,\h
g_{03}=-2l\frac{\delta\cos\theta}{R^2} , \\ \nl
\delta(r)=r^2-2Mr-l^2\h,\h R^2(r)=r^2+l^2 , \ea where $M$ is the mass
and $l$ is the NUT-parameter. Now we have from (\ref{su3}),
(\ref{r.2}) and (\ref{t.2}): \ba\nl \dot t&=&\frac{C_1}{R^2}\Biggl[
C^3\Biggl(\frac{R^4}{\delta}+4l^2\Biggr)-\frac{2l}{\sin^2\theta}\left(
C^0\cos\theta+2C^3 l\right)\Biggr], \\ \nl R^4\dot
r^2&=&\left(C_1C^3\right)^2\left(R^4+4l^2\delta\right)-C_2\delta, \\
\nl R^4\dot\theta^2&=&C_2-\frac{C_1^2}{\sin^2\theta}\Bigl[\left(
C^0\right)^2+\left(2C^3 l\right)^2+4C^0 C^3 l\cos\theta\Bigr],
\\ \nl \dot
\varphi&=&\frac{C_1}{R^2\sin^2\theta}\left(C^0+2C^3
l\cos\theta\right). \ea In the Taub-NUT metric the solution
(\ref{GSC}) is \ba\nl x^0 (\tau,\s^1,\s^2)&=&C^0 f[w(z^1,z^2)] + t_0
\\ \nl
&\pm&\int\limits_{r_0}^{r}dr\Bigl[C^3\left(R^4
\delta^{-1}+4l^2\right)-2l
\sin^{-2}\theta_0\left(C^0\cos\theta_{0}+2C^3
l\right)\Bigr]U^{-1/2}(r) , \\ \nl x^3 (\tau,\s^1,\s^2)&=&C^3
f[w(z^1,z^2)] + \varphi_{0}
\\ \nl
&\pm&\frac{1}{\sin^2\theta_0}\int\limits_{r_0}^{r}dr\left(C^0+2C^3
l\cos\theta_{0}\right)U^{-1/2}(r) ,\\ \nl C_1(\tau - \tau_{0})&=&
\pm\int\limits_{r_0}^{r}drR^2 U^{-1/2}(r) ,\ea where \ba\nl
U(r)=\left(C^3\right)^2\left(R^4-4l^2\delta\cot^2\theta_{0}\right)
-\delta\sin^{-2}\theta_{0}\left[\left(C^0\right)^2+4C^0 C^3
l\cos\theta_{0}\right] . \ea

 Finally, we consider the {\it Kerr space-time} with metric taken
in the form \ba\nl g_{00}&=&-\Biggl(1-\frac{2Mr}{\rho^2}\Biggr)\h,\h
g_{11}=\frac{\rho^2}{\Delta},
\\ \nl
g_{22}&=&\rho^2 \h,\h g_{33}=\Biggl(r^2+a^2+\frac{2Ma^2
r\sin^2\theta}{\rho^2}\Biggr)\sin^2\theta ,
\\ \nl
g_{03}&=&-\frac{2Mar\sin^2\theta}{\rho^2} , \ea where \ba\nl
\rho^2=r^2+a^2\cos^2\theta \h,\h \Delta=r^2-2Mr+a^2 , \ea $M$ is the
mass and $a$ is the angular momentum per unit mass of the Kerr black
hole. With this input in equations (\ref{su3}), (\ref{r.2}) and
(\ref{t.2}), we have: \ba\nl \dot
t&=&\frac{C_1}{\Delta\rho^2}\Bigl[C^3 (r^2+a^2)^2 -C^3 a^2\Delta
\sin^2\theta - 2C^0 M a r\Bigr] ,
\\ \nl
\rho^4\dot r^2&=&C_1^2\Bigl[(C^3)^2 (r^2+a^2)^2
-4C^0 C^3 Mar+(C^0)^2 a^2\Bigr] - C_2 \Delta ,
\\ \nl
\rho^4\dot\theta^2&=&C_2-C_1^2\Biggl(\frac{(C^0)^2}{\sin^2\theta}
+(C^3)^2 a^2 \sin^2\theta\Biggr)
\\ \nl
\dot\varphi&=&\frac{C_1}{\Delta\rho^2}\Biggl( \frac{C^0
\Delta}{\sin^2\theta}+2C^3 Mar-C^0 a^2\Biggr) . \ea In this case, the
exact solution (\ref{GSC}) takes the form: \ba\nl
x^0
(\tau,\s^1,\s^2)&=&C^0 f[w(z^1,z^2)] + t_0
\\ \nl
&\pm&\int\limits_{r_0}^{r}dr\Delta^{-1} \Bigl\{2C^0 Mar-C^3\left[
(r^2+a^2)\rho_{0}^2+2Ma^2 r\sin^2\theta_0\right]\Bigr\}V^{-1/2}(r)
,\\ \label{K} x^3 (\tau,\s^1,\s^2)&=&C^3 f[w(z^1,z^2)] + \varphi_{0}
\\ \nl
&\pm&\frac{1}{\sin^2\theta_0}\int\limits_{r_0}^{r}dr\Delta^{-1}
\left[C^0\left(2Mr-\rho^2_{0}\right)-2C^3
Mar\sin^2\theta_0\right]V^{-1/2}(r) ,\\ \nl C_1(\tau - \tau_{0})&=&
\pm\int\limits_{r_0}^{r}dr\rho_0^2 V^{-1/2}(r) ,\\ \nl
V(r)&=&\left(C^0\right)^2\left(a^2-\Delta\sin^{-2}\theta_{0}\right)
-4C^0 C^3 Mar\\ \nl &+&\left(C^3\right)^2\Bigl[\left(r^2+a^2\right)^2
-a^2\Delta\sin^2\theta_{0}\Bigr] ,\\ \nl
\rho^2_{0}&=&r^2+a^2\cos^2\theta_{0} . \ea

\section{\bf Comments and conclusions}
\hspace{1cm} In the previous section we restrict ourselves to some
particular cases of the generic solution (\ref{GSC}) and do not pay
attention to the existing possibilities for obtaining solutions in
the case $\theta\neq const$ in the considered examples.

Obviously, the examples given in Sec.5 do not exhaust all
possibilities contained in the metric (\ref{gm}) \cite{KSMH}. On the
other hand, in different particular cases of this type of metric,
there exist more general brane solutions. They will be published
elsewhere. Here we only mention that in the gauge $\mu^{k}=const$,
\ba\nl x^{\nu}(\tau,\us)=x^{\nu}(\mu^{k}\tau+\s^k) \ea is an obvious
nontrivial solution of the equations of motion and of the constraints
(\ref{eqx}), (\ref{Tx1}) depending on $D$ arbitrary functions of $p$
variables for the null $p$-brane in arbitrary $D$-dimensional gravity
background.

From the results of the previous sections, it is easy to extract the
corresponding formulas for the null string case simply by putting
$\s^1=\s, \mu^1=\mu, \s^2=\mu^2=0$. For example, our equalities
(\ref{S}) coincide with the ones obtained in \cite{DabLar} for the
null string moving in Schwarzschild space-time after identification:
\ba\nl E=C_1 C^3 ,\h L=C_1 C^0 ,\h L^2+K=C_2 . \ea Moreover, our
solution (\ref{K}) in the case $p=1$, generalizes the solution given
in \cite{PorPap}. The latter corresponds to fixing the arbitrary
function $f(w)$ to a linear one and fixing the gauge to $\mu=0$, i.e.
\ba\nl f[w(\mu^1\tau+\s^1,\mu^2\tau+\s^2)]\mapsto
f(\mu\tau+\s)=\mu\tau+\s ,\h \mbox{with}\h \mu=0 . \ea

In this paper we perform some investigation on the classical dynamics
of the null bosonic branes in curved background. In the second
section we give the action, prove its reparametrization invariance
and give the equations of motion and constraints in an arbitrary
gauge. Then we construct the corresponding Hamiltonian and compute
the constraint algebra. Following \cite{FF}, we obtain the manifest
expressions for the classical BRST charge, the total constraints, the
BRST invariant Hamiltonian and the BRST invariant Lagrangian. All
this gives the possibility for quantization of the null $p$-branes in
a curved space-time. In the fourth section we consider the dynamics
of the null membranes ($p=2$) in a four dimensional, stationary,
axially symmetrical, gravity background. Some exact solutions of the
equations of motion and of the constraints are found there. The next
section is devoted to examples of such solutions in Minkowski, de
Sitter, Schwarzschild, Taub-NUT and Kerr space-times.

\vspace*{1cm}
{\bf Acknowledgments}

The author would like to thank D. Mladenov for useful discussions
and B. Dimitrov for careful reading of the manuscript.


\vspace*{.5cm}


\end{document}